\documentclass[aps,pra,10pt,notitlepage,twocolumn,nofootinbib]{revtex4-2}
\usepackage{amsmath,amsthm,amsfonts,amssymb}
\usepackage{enumerate}
\usepackage{xcolor}
\usepackage{bbm}

\usepackage{tikz}
\usetikzlibrary{calc}
\usetikzlibrary{arrows}
\usetikzlibrary{shadows}
\usetikzlibrary{positioning}
\usetikzlibrary{decorations.pathmorphing,shapes}
\usetikzlibrary{shapes,backgrounds}
\usepackage{mathrsfs}

\newcommand{\ignore}[1]{}

\newcommand{\ie}{{\em i.e.}}

\newtheorem{theorem}{Theorem}

\newcommand{\ZZ}{\mathbb{Z}}
\newcommand{\RR}{\mathbb{R}}
\newcommand{\CC}{\mathbb{C}}

\newcommand{\GG}{\mathcal{G}}
\newcommand{\HH}{\mathcal{H}}
\newcommand{\LL}{\mathcal{L}}
\newcommand{\OO}{\mathcal{O}}
\newcommand{\SbS}{\mathcal{S}}

\DeclareMathOperator{\spn}{span}

\DeclareFontFamily{U}{bbold}{}
\DeclareFontShape{U}{bbold}{m}{n}
 {
  <-5.5> s*[1.069] bbold5
  <5.5-6.5> s*[1.069] bbold6
  <6.5-7.5> s*[1.069] bbold7
  <7.5-8.5> s*[1.069] bbold8
  <8.5-9.5> s*[1.069] bbold9
  <9.5-11> s*[1.069] bbold10
  <11-15> s*[1.069] bbold12
  <15-> s*[1.069] bbold17
 }{}

\newcommand{\norm}[1]{\left\lVert #1 \right\rVert}
\newcommand{\ip}[2]{\langle #1, #2 \rangle}

\renewcommand{\Re}{\mathtt{Re}}
\renewcommand{\Im}{\mathtt{Im}}

\def\oneshot#1{\mathop{\mathrm{#1}}\limits}
\def \limn {\oneshot\lim_{n\rightarrow \infty}}

\makeatletter
\def\bbl@set@language#1{%
  \edef\languagename{%
    \ifnum\escapechar=\expandafter`\string#1\@empty
    \else\string#1\@empty\fi}%
  \@ifundefined{babel@language@alias@\languagename}{}{%
    \edef\languagename{\@nameuse{babel@language@alias@\languagename}}%
  }%
  \select@language{\languagename}%
  \expandafter\ifx\csname date\languagename\endcsname\relax\else
    \if@filesw
      \protected@write\@auxout{}{\string\select@language{\languagename}}%
      \bbl@for\bbl@tempa\BabelContentsFiles{%
        \addtocontents{\bbl@tempa}{\xstring\select@language{\languagename}}}%
      \bbl@usehooks{write}{}%
    \fi
  \fi}

\begin{document}
\title{No Infinite Tail Beats Optimal Spatial Search}
\author{Weichen Xie}
\email{xiew@clarkson.edu}
\affiliation{Department of Mathematics, Clarkson University, Potsdam, New York, USA 13699-5815}
\author{Christino Tamon}
\email{tino@clarkson.edu}
\affiliation{Department of Computer Science, Clarkson University, Potsdam, New York, USA 13699-5815}
\date{\today}
\begin{abstract}
Farhi and Gutmann ({\em Physical Review A}, {\bf 57}(4):2403, 1998) proved that a continuous-time analogue 
of Grover search (also called spatial search) is optimal on the complete graphs. 
We extend this result by showing that spatial search remains optimal in a complete graph 
even in the presence of an infinitely long path (or tail). If we view the latter as an external quantum system
that has a limited but nontrivial interaction with our finite quantum system,
this suggests that spatial search is robust against a coherent infinite one-dimensional probe.
Moreover, we show that the search algorithm is {\em oblivious} in that it does not need to
know whether the tail is present or not, and if so, where it is attached to.
\end{abstract}
\maketitle

\section{Introduction}

The celebrated quantum search algorithm of Grover \cite{g97} provides a provable quadratic speedup over any classical algorithm. 
Shortly thereafter, Farhi and Gutmann proposed an analogue analog of Grover's algorithm \cite{fg98}. 
They defined the Grover search algorithm via a continuous-time quantum walk on a complete graph where the oracle or target vertex 
is marked by a suitably weighted self-loop. Remarkably, the Farhi-Gutmann algorithm achieved perfect fidelity 
on complete graphs of any size. In contrast, this property does not hold for Grover search (viewed as a discrete-time quantum walk)
as it is inherently a bounded-error probabilistic algorithm.

This continuous-time search problem was later generalized by Childs and Goldstone \cite{cg04} to arbitrary finite graphs 
where it is known as the \emph{spatial search} problem.
A collection of different families of finite graphs had been studied in this context; 
for example, see \cite{cg04,jmw14,w16,gw18,cnr20,tsp22}. 
But, to date, spatial search has not been studied on {\em infinite} graphs as it seems that the quantum walk 
will escape or diffuse to infinity before having a chance to localize on the marked vertex (or oracle).
The goal of this work is to disabuse ourselves of this highly plausible intuition. 

We consider infinite graphs which are obtained by attaching an infinite path (or tail) to a finite graph.
This family of graphs with tails was explored by Golinskii \cite{g16}. 
In this work, we view the finite graph as our operational quantum system for performing quantum search 
and the tail as an external (possibly, adversarial) infinite-dimensional quantum system which interacts 
with our finite sytem in a coherent manner. 

The main question we explore in this work is: {\em can spatial search still be performed optimally 
in the presence of an infinite-dimensional probe?} We provide a positive answer to this question 
for complete graphs. This extends the result of Farhi-Gutmann \cite{fg98} to the infinite setting.
Moreover, the quantum search algorithm is {\em oblivious} as it does not need to know 
whether the infinite-dimensional probe is present (or not) and where it is attached to (if present).
Since we give our adversary the benefit of an infinite-dimensional quantum system, this serves only
to strengthen the result. 

Our technique relies on the theory of Jacobi operators (see \cite{g16,ds06}).
The main idea is to decompose the adjacency operator $A$ of our infinite graph using
two pairwise orthogonal invariant subspaces (see Golinskii \cite{g16}, Theorem 1.2) 
where the first one is finite-dimensional while the second one is infinite-dimensional. 
The next crucial observation is that spatial search takes place in the infinite-dimensional invariant subspace of $A$.
Moreover, the action of $A$ in this infinite-dimensional invariant subspace is given by a finite rank Jacobi matrix $J$. 
Finally, we show that the initial and target states of the spatial search are
nearly confined in a two-dimensional subspace spanned by two {\em bound} states of $J$.
This yields the claimed spatial search result.
To the best of our knowledge, this is the first result which explores optimal spatial search on infinite graphs.

As outlined above, our argument is the standard argument for showing optimal spatial search in the finite setting
(see \cite{cnao16,cnr20,cgtx22}).
Namely, we show that the initial and target states are spanned by two distinct eigenvectors of the perturbed
adjacency matrix. A contribution of this work is to show that this argument holds in the infinite setting
via bound states of the reduced adjacency operator. We believe that this argument might be useful in other
settings. 

Following a brief discussion of basic notation and terminology in Section \ref{section:preliminary}, 
we prove our main results in Sections \ref{section:lollipop1} and \ref{section:lollipop2}.
Then, we justify the optimality of our search algorithm in Section \ref{section:lower-bound}.
Finally, we conclude with some open questions in Section \ref{section:conclusions}.

\section{Preliminaries}\label{section:preliminary}

We introduce the basic notation and terminology that we will use throughout. 
The set of all positive integers is denoted $\mathbb{Z}^+$
and the set of all complex numbers with unit modulus is denoted $U(1)$.
For vectors $x,y$ where $x = \zeta y$, for some $\zeta \in U(1)$, we write $x \equiv y$. 
We adopt standard asymptotic notation:
$o(f_n)$ denotes any function $g_n$ so that $g_n/f_n\to 0$,
$\OO(f_n)$ denotes functions $g_n$ for which $g_n/f_n$ is bounded from above by a constant,
and
$\Omega(f_n)$ denotes functions $g_n$ where $g_n/f_n$ is bounded from below by a constant,
where in each case $n\to\infty$; see \cite{t39}. 
In our case, the asymptotic parameter $n$ corresponds to the size of a finite graph.

\paragraph{Graphs and operators.}
We study undirected and connected graphs $G=(V,E)$ with vertex set $V$ and edge set $E$, respectively. 
The adjacency matrix $A$ of $G$ is a symmetric matrix whose $(i,j)$ entry is $1$ if $(i,j)\in E$ and $0$ otherwise.
For a vertex $u$, let $N(u) = \{v \in V : (u,v) \in E\}$ denote the set of neighbors of $u$.
The degree of vertex $u$, denoted $\deg(u)$, is the cardinality of $N(u)$.
The complete graph (or clique) on $n$ vertices is denoted $K_n$. 
A rooted graph $(G,r)$ is a graph $G$ with a distinguished vertex $r$ which we call the root.
See \cite{gr01} for further background on algebraic graph theory.

We allow countably infinite graphs, in which case, $V = \ZZ^+$ (see \cite{mw89}).
For example, the infinite path $P_\infty$ has edges which are consecutive positive integers;
its adjacency matrix is known as the free Jacobi matrix (see \cite{g16}).
Related to these graphs, we associate a complex separable Hilbert space $\HH = \ell^2(V)$ equipped 
with the inner product $\ip{x}{y}=\sum_{u\in V}\overline{x_u}y_u$, for vectors $x,y\in \HH$.
For $\ell^2(V)$, a standard basis is $\{e_u : u\in V\}$, where $e_u$ is the unit vector 
corresponding to vertex $u$.
An infinite graph $\GG$ is \emph{locally finite} if $\deg{(u)}<\infty$ for all $u\in V$. 
For such a graph $\GG$, the adjacency operator $A$ is a linear operator that maps the standard basis vector $e_v$ 
to the vector associated with the neighboring vertices $N(v)$; 
that is, $Ae_v=\sum_{u\in N(v)}a_{u,v}e_u$, or simply $a_{u,v}=\ip{e_u}{Ae_v}$.
Let $\deg{(\GG)}=\sup\{\deg{(u)} : u \in V\}$.
If $\deg(\GG) < \infty$, then the adjacency operator $A$ is a bounded self-adjoint operator (see \cite{m82}).

The \emph{spectrum} of a linear operator $A$ is the set $\sigma(A)$ of all complex numbers $\lambda$ where $\lambda I-A$ is not invertible. 
For a bounded and self-adjoint operator $A$, its spectrum can be classified further into the \emph{point} spectrum $\sigma_p(A)$ 
and the \emph{continuous} spectrum $\sigma_c(A)$. 
The point spectrum consists of all eigenvalues $\lambda \in \CC$ of $A$ such that $Ax=\lambda x$ for some nonzero $x \in \ell^2(V)$.
On the contrary, the values in $\sigma_c(A)$ are not eigenvalues of $A$ and have no corresponding eigenvectors in $\ell^2(V)$.

The spectral theorem (see \cite{ag13,r91}) states that
for a bounded self-adjoint operator $A$ on a complex Hilbert space $\HH$, 
there exists a unique resolution of the identity $E$ on the Borel subsets of $\sigma(A)$ so that $A = \int_{\sigma(A)}\lambda \ dE(\lambda)$.
Moreover, if $f$ is a bounded Borel function on $\sigma(A)$, then $f(A) = \int_{\sigma(A)} f(\lambda) \ dE(\lambda)$.
We also use a decomposition induced by invariant subspaces (see \cite{ag13}, Theorem 3, section 40)
which states that
if $W_k$ $(k=1,2,\cdots,m)$ are pairwise orthogonal invariant subspaces of $A$, that is, $\HH=\bigoplus_{k=1}^{m} W_k$ 
and $AW_k\subset W_k$, for each $k$, then $A = \sum_{k=1}^m A_k P_k$, where $P_k$ is the projection on $W_k$ and $A_k$ 
is the restriction of $A$ to $W_k$.

\paragraph{Spatial Search.}
A continuous-time quantum walk on an infinite graph $\GG$ with bounded self-adjoint adjacency operator $A$ 
is given by the unitary operator $e^{-itA}$ (acting on the Hilbert space $\ell^2(V)$).
Our focus is on infinite graphs obtained from a finite connected rooted graph $(G,r)$ by attaching 
an infinite path $P_\infty$ at the root vertex $r$; denote the resulting infinite graph as $\GG = G(r,P_\infty)$. 
As we use $\{1,2,\ldots,n\}$ to label the vertices of a finite rooted graph of order $n$, we take the liberty to
designate the last vertex $n$ as the root (without loss of generality). 
These are the graphs with tails studied by Golinskii \cite{g16}.

We say the infinite graph $\GG=G_n(P_\infty)$ has {\em optimal spatial search} (adopting \cite{cnr20})
if there is a real $\gamma > 0$ so that for each vertex $w$ of $G_n$, 
a continuous-time quantum walk on $\GG$ with a self-loop on $w$ of weight $\gamma$
will unitarily map the principal eigenvector $z_1$ of $G_n$ to the unit vector $e_w$ with constant fidelity 
in time $t = \OO(1/\epsilon_1)$, where $\epsilon_1 = |\ip{e_w}{z_1}|$. That is,
\begin{align*}
    \left\vert \ip{e_w}{e^{-it(A+\gamma P_{w})} z_1} \right\vert = \Omega(1), 
\end{align*}
where $A$ is the adjacency operator of $\GG$ and $P_w$ is the projection onto the subspace spanned by $e_w$.
It is customary to assume $\epsilon_1 = o(1)$ as otherwise we already have a constant overlap between the target state $e_w$ 
and the initial state $z_1$.

\section{Infinite lollipop is optimal}\label{section:lollipop1}

\begin{figure}[h]
    \centering
    \includegraphics[width=0.45\textwidth]{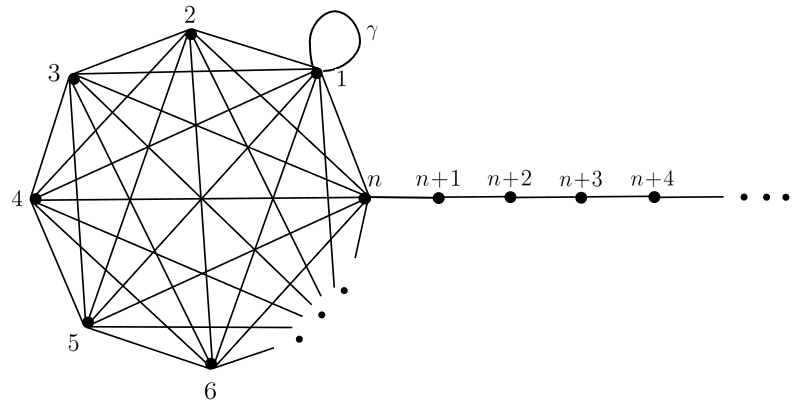}
    \caption{Optimal spatial search on the infinite lollipop with the oracle hiding at vertex $1$.}
    \label{fig:lollipop}
\end{figure}

For $n \ge 2$, consider the infinite {\em lollipop} graph $\LL_n = K_n(P_\infty)$.
The vertices of the infinite path $P_{\infty}$ are labelled with $n+1,n+2,\ldots$.
To this lollipop graph $\LL_n$, we place a self-loop of weight $\gamma$ at vertex $1$ (the oracle or target vertex)
and denote the resulting infinite graph as $\LL_n(\gamma)$. Please see Figure \ref{fig:lollipop}.

The adjacency operator $H$ of $\LL_n(\gamma)$ is obviously a bounded self-adjoint linear operator on $l^2(\ZZ^+)$. 
It can be written as an infinite dimensional matrix under the standard basis $\{e_j\}_{j=1}^{\infty}$ as
\begin{align*}
    H =
    \begin{pmatrix}
        \gamma & 1 & \cdots & 1 & & & & \\
        1 & 0 & \cdots & 1 & & & & \\
        \vdots & \vdots & \ddots & \vdots & & & & \\
        1 & 1 & \cdots & 0 & 1 & & & & \\
         & & & 1 & \ 0 & 1 & & \\
         & & & & \ 1 & 0 & 1 & \\
         & & & & & \ddots & \ddots & \ddots
    \end{pmatrix}.
\end{align*}

Our goal is to prove the following infinite analogue of the Farhi-Gutmann result.

\begin{theorem} \label{thm:lollipop}
For $\gamma = n + \OO(1)$, we have 
\begin{align*}
|\ip{e_1}{e^{-itH} z_1}| = \Omega(1),
\end{align*}
for $t = \pi/2\sqrt{n}$, 
where $z_1 = n^{-1/2}\sum_{j=1}^{n} e_j$.
\end{theorem}

By a theorem of Golinskii (\cite{g16}, Theorem 1.2), using a change of basis, $H$ can be written in a block diagonal form. 
This new basis $\{\tilde{e}_k\}_{k=1}^{\infty}$ is defined as follows. 
Let
\begin{align}
    \Tilde{e}_k=e_k,\  \ k=n,n+1,\cdots
\end{align}
and
\begin{align}
    \Tilde{e}_{n-1}=\frac{1}{\sqrt{n-1}}\sum_{j=1}^{n-1}e_j.
\end{align}
The next orthogonal basis vector is then given by
\begin{align*}
    \Tilde{e}_{n-2}' &= H\Tilde{e}_{n-1}-\ip{\Tilde{e}_{n-1}}{H\Tilde{e}_{n-1}}\Tilde{e}_{n-1}-\ip{\Tilde{e}_{n}}{H\Tilde{e}_{n-1}}\Tilde{e}_{n}.
\end{align*}
which, after normalization, yields
\begin{align}
    \Tilde{e}_{n-2}=\frac{1}{\sqrt{(n-2)(n-1)}}
	((n-1)e_1 - \Tilde{e}_{n-1}).
\end{align}
We take the remaining basis vectors to be the non-principal columns of the Fourier matrix of order $n-2$. 
In particular, the basis vector $\tilde{e}_{k}$ is defined as
\begin{align*}
\tilde{e}_{k} = \frac{1}{\sqrt{n-2}}\sum_{j=1}^{n-2} e^{\frac{2\pi i(j-1)k}{(n-2)}}e_j
\ \ \
(k=1,\ldots,n-3).
\end{align*}
Since $H\Tilde{e}_{n-2}=\left(\gamma\frac{n-2}{n-1}-1\right)\Tilde{e}_{n-2}+\gamma\frac{\sqrt{n-2}}{n-1}\Tilde{e}_{n-1}$, 
the subspace $\SbS=\spn\{\Tilde{e}_{n-2},\Tilde{e}_{n-1},\cdots\}$ is $H$-invariant. 
Under the new basis, the adjacency operator $H$ becomes
\begin{align*}
	H = 
	\begin{pmatrix}
	-I_{n-2} & O \\
	O & \widehat{H}
	\end{pmatrix}
\end{align*}
As will be clear soon, it suffices for us to restrict our focus on the operator $\widehat{H}$, 
which is the operator $H$ restricted to the subspace $\SbS = \spn\{\Tilde{e}_{n-2},\Tilde{e}_{n-1},\cdots\}$. 
This is because the initial state $z_1$ and the target state $e_1$ of our spatial search problem 
both have non-negligible overlap with $\SbS$.

It follows from the preceding analysis that the operator $H$ under the basis $\{\Tilde{e}_{n-2},\Tilde{e}_{n-1},\ldots\}$
is given by
\begin{align}
    \widehat{H} = 
    \begin{array}{rl}
    \begin{matrix}
        {\scriptstyle \Tilde{e}_{n-2}}\\
        {\scriptstyle \Tilde{e}_{n-1}}\\
        {\scriptstyle \Tilde{e}_{n}}\\
        {\scriptstyle \Tilde{e}_{n+1}}\\
        \vdots
    \end{matrix}
    \!\!\!\!\!
         &  
         \begin{pmatrix}
            \gamma\frac{n-2}{n-1}\!-\!1 & \gamma\frac{\sqrt{n-2}}{n-1} & & & & & \\
            \gamma\frac{\sqrt{n-2}}{n-1} & n\!\!-\!\!2\!+\!\frac{\gamma}{n-1} & \sqrt{n\!\!-\!\!1} & & & & \\
             & \sqrt{n\!\!-\!\!1} & 0 & 1 & & & \\
             & & 1 & 0 & \ \ 1 & & \\
             & & & \ddots & \ddots & \ddots
        \end{pmatrix}.
    \end{array}
\end{align}
This symmetric tridiagonal matrix is an \emph{eventually-free} or {\em finite rank} Jacobi matrix (see \cite{g16}) 
whose full spectrum can be computed via the so-called \emph{Jost} solution (see \cite{ds06}, Appendix).
The Jost solution is a vector $y(x)=(y_1(x),y_2(x),\cdots)^T$ that satisfies the eigen-equation of 
$\widehat{H}$ with eigenvalue of the form $x+\frac{1}{x}$, namely,
\begin{align} \label{eqn:eigen}
    \widehat{H}y(x) = \left(x+\frac{1}{x}\right)y(x),
\end{align}
and also the degree condition
\begin{align*}
    \lim_{k\to\infty}x^{-k}y_k(x)=1.
\end{align*}
Given the special form of $\widehat{H}$, we can set 
\begin{align}
    y_k(x) = x^k,\ \ k=3,4,\cdots,
\end{align}
and use \eqref{eqn:eigen}, to get the Jost polynomial
\begin{align}
    y_0(x) &= \frac{1}{\gamma}\sqrt{\frac{n-1}{n-2}}\cdot\left[
    (2-n)x^4 + [(n-3)\gamma+4-2n)]x^3 \right.\nonumber\\
    &+ \left.[(n-3)\gamma+5-2n]x^2 + [3-n-\gamma]\,x + 1\right]. 
\end{align}
In order to compute the eigenvalues of $\widehat{H}$, we need the following spectral theorem for
finite rank Jacobi operators. 
We will only need information about the point spectrum as will be clear soon.

\begin{theorem} (\cite{g16}, p8)
    Let $J$ be an eventually-free Jacobi matrix and $y_0(x)$ be its Jost function. 
	Then all roots of $y_0(x)$ in the complex unit disk are real and simple, $y_0(0)\not=0$. 
	A real number $\lambda_j$ is an eigenvalue of $J$ \emph{if and only if} 
    \begin{align*}
        \lambda_j=x_j+\frac{1}{x_j},\quad x_j\in(-1,1),\quad y_0(x_j)=0.
    \end{align*}
\end{theorem}

By choosing $\gamma=n+\OO(1)$, the roots of $y_0(x)$ in the complex unit disk can be approximated consecutively.
First note that there are four real roots for $y_0(x)$ and two of them lie in the unit disk as indicated by the following table:
\begin{table}[h]
    \centering
    \begin{tabular}{c|c|c|c|c|c}
        $x$   & $-\infty$   & $-1$  & ${\scriptstyle \frac{1}{n}}$              & 
        $1$           & $+\infty$ \\
        \hline
        ${\gamma}{\scriptstyle \sqrt{\frac{n-2}{n-1}}}\cdot y_0$ & $-\infty$ & $\gamma\!+\!1\!>\!0$ & ${\scriptstyle -\frac{1}{n}+\OO(\frac{1}{n^{\scriptscriptstyle 2}})}\!<\!0$  & $2n\gamma\!+\!{\scriptstyle \OO(n)}\!>\!0$ & $-\infty$
    \end{tabular}
    \caption{For $\gamma=n+\OO(1)$ and sufficiently large $n$, there are $4$ sign changes of $y_0(x)$. }
    \label{tab:root_y0}
\end{table}

Denote the two roots within the unit disk as $x_{\pm}=\frac{1}{n}+\delta_{\pm}$. Hence,
\begin{align*}
    0 &= {\gamma}\sqrt{\frac{n-2}{n-1}}\cdot y_0(x_{\pm})= -\frac{1}{n}+n^2\delta_{\pm}^2+o\left(\frac{1}{n}\right)+o(n^2\delta_{\pm}^2).
\end{align*}
In order for the left-hand side to attain $0$ exactly, for all $n$, at least the two highest order terms on the right-hand side 
should cancel perfectly; that is, $o(\frac{1}{n})=-\frac{1}{n}+n^2\delta_{\pm}^2$ which implies
\begin{align}
    x_{\pm}=\frac{1}{n}\pm\frac{1}{n^{3/2}}+o\left(\frac{1}{n^{3/2}}\right).
\end{align}
Therefore, the two distinct eigenvalues of $\widehat{H}$ are given by
\begin{align}
    \lambda_{\pm} = n\pm\sqrt{n}+\OO(1).
\end{align}
The corresponding eigenvectors (or bound states) are given by 
$y_{\pm} = (y_1(x_\mp),y_2(x_\mp), \ldots)^T$
where the entries are defined by the Jost polynomials
\begin{align}
    y_1(x_{\mp}) 
    &=
    \pm\frac{1}{\gamma}\sqrt{\frac{n-1}{n-2}}\cdot\frac{1}{n^{3/2}}+\OO\left(\frac{1}{n^3}\right),\\
    y_2(x_{\mp}) &= \frac{1}{\sqrt{n-1}}\cdot\frac{1}{n^2}\mp\frac{1}{\sqrt{n-1}}\frac{2}{n^{5/2}}+\OO\left(\frac{1}{n^{7/2}}\right),\\
    y_k(x_{\mp}) &= \frac{1}{n^k}\mp \frac{k}{n^{(2k+1)/2}}+\OO\left(\frac{1}{n^{k+1}}\right),\quad k=3,4,\cdots.
\end{align}
Now, we are ready to prove Theorem \ref{thm:lollipop}. 

First, note that both
$e_1=\sqrt{\frac{1}{n-1}}\Tilde{e}_{n-2}+\sqrt{\frac{n-2}{n-1}}\Tilde{e}_{n-1}$ and $z_1=\Tilde{e}_{n-1}$ are completely 
in the invariant subspace $\SbS$. Thus, we can restrict our unitary evolution to $\SbS$.
Moreover, as $e_1$ overlaps almost completely with $\Tilde{e}_{n-1}$, it suffices to consider $\Tilde{e}_{n-1}$ as the target state.
Hence, the fidelity can be further approximated as
\begin{align*}
    \left\vert \ip{e_1}{e^{-itH}z_1} \right\vert
    \ge \sqrt{\frac{n-2}{n-1}}\left\vert \ip{\overline{e}_1}{e^{-it\widehat{H}}\overline{e}_2} \right\vert+\OO\left(\frac{1}{n^{1/2}}\right),
\end{align*}
where $\overline{e}_k$ is a unit vector which is the $k$-th basis vector 
for the invariant subspace $\SbS=\spn\{\Tilde{e}_{n-2},\Tilde{e}_{n-1},\cdots\}$. 

Notice that 
\begin{align*}
\limn{\frac{\ip{\overline{e}_k}{y_+}\ip{y_+}{\overline{e}_k}}{\ip{y_+}{y_+}}
+ \frac{\ip{\overline{e}_k}{y_-}\ip{y_-}{\overline{e}_k}}{\ip{y_-}{y_-}}} = 1,
\ \
k=1,2,
\end{align*}
that is, both the initial and the target states lie in the two-dimensional invariant subspace spanned 
by the eigenvectors $y_+$ and $y_-$.
Thus, it again suffices to consider the fidelity in this subspace.

Straightforward calculation shows that when $\gamma = n + \OO(1)$, the fidelity satisfies 
$|\ip{e_1}{e^{-itH}z_1}| = 1+o(1)$ for time $t=\frac{\pi}{2\sqrt{n}}$.
If we normalize the adjacency matrix of $K_n$, we obtain $\OO(\sqrt{n})$ time (matching Grover search).

\section{Oracle at the edge of infinity} \label{section:lollipop2}

\begin{figure}[h!!!]
    \centering
    \includegraphics[width=0.45\textwidth]{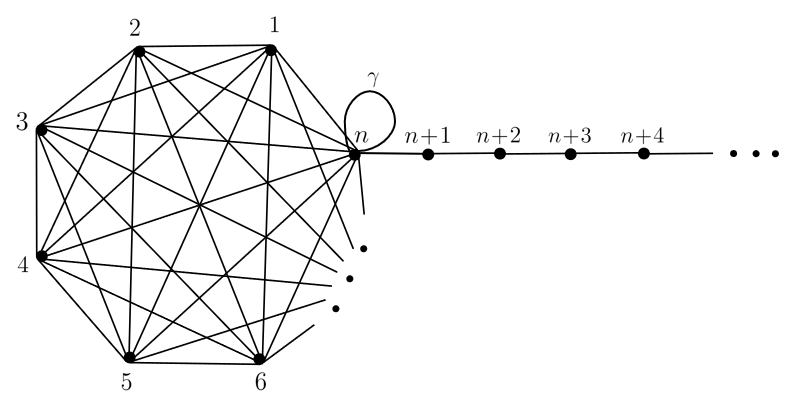}
    \caption{Optimal spatial search still occurs on the infinite lollipop even if the oracle sits at gateway vertex $n$.}
    \label{fig:lollipop2}
\end{figure}

In this section, we show that even when the oracle is placed at the attachment vertex 
of the infinite path, spatial search remains optimal on $\LL_n$. 
See Figure \ref{fig:lollipop2}.

Together with Section \ref{section:lollipop1}, this will show that the search algorithm 
is {\em oblivious} as it does not need to know if the external probe (tail) is present or not.
This is because the two cases surprisingly require the same asymptotic time for optimal spatial search.
The claim follows from a similar analysis as before.

The adjacency operator of $\LL_n(\gamma)$ is given as
\begin{align*}
    H =
    \begin{pmatrix}
        0 & 1 & \cdots & 1 & & & & \\
        1 & 0 & \cdots & 1 & & & & \\
        \vdots & \vdots & \ddots & \vdots & & & & \\
        1 & 1 & \cdots & \gamma & 1 & & & & \\
         & & & 1 & \ 0 & 1 & & \\
         & & & & \ 1 & 0 & 1 & \\
         & & & & & \ddots & \ddots & \ddots
    \end{pmatrix}.
\end{align*}
First, we restrict our focus to the invariant subspace $\SbS=\spn\{\Tilde{e}_{n-1},\Tilde{e}_n,\cdots\}$ 
since $e_n=\Tilde{e}_n\in\SbS$ and $z_1=\sqrt{\frac{n-1}{n}}\Tilde{e}_{n-1}+\sqrt{\frac{1}{n}}\Tilde{e}_n\in\SbS$, 
where 
$\Tilde{e}_{n-1}=\frac{1}{\sqrt{n-1}}\sum_{j=1}^{n-1} e_{j}$
and $\Tilde{e}_k=e_k$, $k\ge n$.
Under the new basis that spans $\SbS$, the operator $H$ is a rank-$2$ Jacobi matrix
\begin{align}
    \widehat{H} = 
    \begin{array}{rl}
    \begin{matrix}
        {\scriptstyle \Tilde{e}_{n-1}}\\
        {\scriptstyle \Tilde{e}_{n}}\\
        {\scriptstyle \Tilde{e}_{n+1}}\\
        \vdots
    \end{matrix}
    \!\!\!\!\!
         &  
         \begin{pmatrix}
            n\!\!-\!\!2\! & \sqrt{n\!\!-\!\!1} & & & & \\
            \sqrt{n\!\!-\!\!1} & \gamma & 1 & & & \\
            & 1 & 0 & \ \ 1 & & \\
            & & \ddots & \ddots & \ddots
        \end{pmatrix}.
    \end{array}
\end{align}
Hence, the following Jost polynomials are obtained:
\begin{align}
    y_0(x) &= \frac{1}{\sqrt{n-1}}\cdot[-\gamma x^3+(n-2)(\gamma-1)x^2\nonumber\\
    &\ \ \ \ \ \ \ \ \ \ \ \ \ \ \ \ \ +(2-n-\gamma)x+1],\\
    y_1(x) &=\frac{1}{\sqrt{n-1}}\cdot(x-\gamma x^2),\\
    y_k(x) &=x^k,\ \ k=2,3,\cdots.
\end{align}
By choosing $\gamma=n+\OO(1)$, we can compute the two distinct roots for $y_0$ that is within the interval $[-1,1]$
\begin{align}
    x_{\pm}=\frac{1}{n}\pm\frac{1}{n^{3/2}}+o\left(\frac{1}{n^{3/2}}\right),
\end{align}
which yields the corresponding eigenvalues for $\widehat{H}$
\begin{align}
    \lambda_{\pm}=n\pm\sqrt{n}+\OO(1).
\end{align}
As shown previously, the corresponding eigenvectors are defined by the values of Jost polynomials:
\begin{align}
    y_1(x_{\mp}) &= \pm\frac{1}{\sqrt{n-1}}\cdot\frac{1}{n^{3/2}}+\OO\left(\frac{1}{n^{5/2}}\right)\\
    y_2(x_{\mp}) &= \frac{1}{n^2}\mp\frac{2}{n^{5/2}}+\OO\left(\frac{1}{n^3}\right)\\
    y_k(x_{\mp}) &= \frac{1}{n^k}\mp\frac{k}{n^{(2k+1)/2}}+\OO\left(\frac{1}{n^{k+1}}\right),\ \ k=3,4,\cdots.
\end{align}
Following the same argument, we can see that for time $t=\frac{\pi}{2\sqrt{n}}$, the fidelity satisfies $|\ip{e_n}{e^{-itH}z_1}|=1+o(1)$.

\section{Optimality}\label{section:lower-bound}
\newcommand{\CQ}{\mathcal{Q}}
\newcommand{\boo}{\mathfrak{B}}
We show that the time bound obtained in Theorem \ref{thm:lollipop} (and in Section \ref{section:lollipop2}) is optimal.
To this end, we generalize an argument of Farhi and Gutmann \cite{fg98} 
to a class of infinite graphs with tails.

For a finite graph $G_n$ on $n$ vertices, take the cone $\widehat{G}_n = K_1 + G_n$ which is
obtained by adding a new vertex (called the conical vertex) and connecting it to all vertices of $G_n$.
Then, attach a tail to the conical vertex and denote this infinite graph as $\widehat{G}_n(P_\infty)$.
Notice that we recover the infinite lollipop when $G_n$ is a clique.

\begin{theorem} \label{thm:lower-bound}
Let $G_n$ be a $(n,d)$-regular graph, where $d = \omega(\sqrt{n})$ for some $\delta > 0$,
and let $\HH_0$ be the adjacency operator of $\widehat{G}_n(P_\infty)$.
Let $z_1 \in \ell^2$ being natural embedding of the principal eigenvector of $G_n$.
Suppose there is a time $t_0$ so that for some $\gamma \in \RR$ and $\zeta \in U(1)$, 
and for a vertex $w$ of $G_n$, we have
\[
    \norm{e^{-it_0(\HH_0 + \gamma P_{w})}z_1 - \zeta e_{w}}^2 = o(1).
\]
Then, $\gamma t_0 = \Omega(1/\epsilon_1)$, where $\epsilon_1 = |\ip{e_w}{z_1}|$,
provided $1/\gamma\epsilon_1 \in o(1)$.
\end{theorem}

The largest eigenvalue in $\sigma_p(\HH_0)$ has a {\em unique bound eigenstate} 
$\beta_1$ which satisfies $\norm{\beta_1 - z_1} = o(1)$. 
This can be shown using similar techniques as in previous sections.
\ignore{
By a result of Golinskii (\cite{g16}, Theorem 1.2), $\HH_0$ can be decomposed as
\begin{align*}
\HH_0 = 
	\begin{pmatrix} 
	A_{n-1} & O \\ 
	O & \widetilde{\HH} 
	\end{pmatrix},
	\ \mbox{ where } \
\widetilde{\HH} = 
	\begin{pmatrix}
	d 		 & \sqrt{n} &        & 			& \\
	\sqrt{n} & 0        & 1      & 			& \\
	         & 1        & 0      & 1 		& \\
	         &          & \ddots & \ddots	& \ddots
	\end{pmatrix}.
\end{align*}
Recall that the first basis vector in the above representation of $\widetilde{\HH}$ is given by
$(z_1,0,0,\ldots)$ where $z_1$ is the all-one vector of order $n$.
The Jost polynomial $y_0(x)$ satisfies $\sqrt{n}y_0(x) = (1-n)x^2 - dx + 1$ whose roots are
\begin{align*}
x_\pm = \frac{d \pm \sqrt{d^2 + 4(n-1)}}{2(1-n)}.
\end{align*}
For $d = \Omega(n^{0.5+\delta})$, $\hat{x} \sim 1/n^{0.5+\delta}$ is the relevant root in $(-1,1)$.
The corresponding eigenvector is $\beta_1 = (y_1(\hat{x}), y_2(\hat{x}), \ldots)$ where
$y_1(\hat{x}) = \hat{x}/\sqrt{n}$, $y_2(\hat{x}) = \hat{x}^2$, 
and so on. 
Therefore, $\beta_1 = (1, 1/n^{0.5+\delta}, 1/n^{0.5+2\delta}, \ldots)$ (up to normalization).
This shows that $\norm{\beta_1 - z_1} = o(1)$.
}

We call a time-dependent state $\psi(t) \in \ell^2(\ZZ^+)$ {\em exponentially decaying} if there is 
a positive integer $M$ so that for all $m \ge M$ we have $|\ip{e_m}{\psi(t)}| = \OO(\kappa^{m})$, 
for some $\kappa < 1$, for all $t$.

\begin{proof} (of Theorem \ref{thm:lower-bound})
Let $\HH_w = \HH_0 + \gamma P_w$.
It suffices to show the claim under the assumption 
$\norm{e^{-it\HH_w}\beta_1 - \zeta e_w}^2 = o(1)$, 
since $\norm{\beta_1 - z_1} = o(1)$ and by using the triangle inequality for squared norm 
(see \cite{s19}, eq. 18.5).

Following \cite{fg98}, we compare two Schr\"{o}dinger evolutions given by
\begin{align*}
\psi_{w}'(t) &= -i\HH_w\psi_w(t), & \psi_w(0) = \beta_1, \\
\psi_{0}'(t) &= -i\HH_0\psi_0(t), & \psi_0(0) = \beta_1.
\end{align*}
Note $\psi_0(t) \equiv \beta_1$, for all $t$, as $\beta_1$ is an eigenstate of $\HH_0$.

For simplicity, we assume that spatial search achieves perfect fidelity, namely, $\psi_w(t_0) \equiv e_w$.
The general case is handled using triangle inequality for squared norm.

The key quantity is $M(t) := \norm{\psi_w(t) - \psi_0(t)}^2$.
First, notice that
\begin{align} \label{eqn:lbound}
M(t_0) &= 2(1-\Re\ip{e_w}{\beta_1}) \ge 2(1-\epsilon_1).
\end{align}
Furthermore, we have
$M'(t) = -2\Re \ip{\psi_w(t)}{\psi_0(t)}'$. 
Given that the inner product is an infinite series, the existence of its derivative requires uniform convergence.
As $\psi_0(t) \equiv \beta_1$ is exponentially decaying (by virtue of being a Jost solution), 
we have $|\overline{(\psi_{w}(t))_m}(\beta_1)_m| \le |(\beta_1)_m|$ and
\begin{align*}
\sum_{m=1}^{\infty} |(\beta_1)_m| 
	\le C_M + \sum_{m \ge M} \kappa^m 
	= C_M + \frac{\kappa^M}{1-\kappa} < \infty
\end{align*}
for a constant $C_M$. 
Thus, $\ip{\psi_w(t)}{\beta_1}$ is uniformly convergent (see Titchmarsh \cite{t39}, 1.11).

Since $\HH_w$ is a finite-rank Jacobi matrix and $\psi'_w(t) = -i\HH_w\psi_w(t)$,
we see that $\ip{\psi'_w(t)}{\psi_0(t)}$ is also uniformly convergent.
This allows us to take the derivative of $\ip{\psi_w(t)}{\psi_0(t)}$ by termwise differentiation
(see Titchmarsh \cite{t39}, 1.72), 
\ie,
\begin{align*}
\ip{\psi_w(t)}{\psi_0(t)}' 
	&\!=\! \ip{\psi_w(t)}{\psi'_0(t)} \!+\! \ip{\psi'_w(t)}{\psi_0(t)} \\
	&\!=\! -i\ip{\psi_w(t)}{\HH_0\psi_0(t)} \!+\! i\ip{\HH_w\psi_w(t)}{\psi_0(t)}.
\end{align*}
So, we obtain
\begin{align*}
M'(t) = 2\gamma\Im\ip{P_w\psi_w(t)}{\psi_0(t)} \le 2\gamma\norm{P_w\psi_0(t)}.
\end{align*}
Thus, $M'(t) \le 2\gamma\epsilon_1$, which further implies
\begin{align*}
M(t_0) = \int_{0}^{t_0} M'(t) dt \le 2\gamma\epsilon_1 t_0.
\end{align*}
By combining the lower and upper bounds on $M(t_0)$, we get
$\gamma t_0 \ge (1-\epsilon_1)/\epsilon_1 = \Omega(1/\epsilon_1)$
as $\epsilon_1 = o(1)$.
\end{proof}

Theorem \ref{thm:lower-bound} justifies the optimal time $1/\epsilon_1$ used to define spatial search.
For the infinite lollipop, we have $\epsilon_1 = 1/\sqrt{n}$, and thus, $\gamma t_0 = \Omega(\sqrt{n})$. 
Since the search algorithm uses $\gamma = n+\OO(1)$, we get $t_0 = \Omega(1/\sqrt{n})$, which matches 
the bound achieved by Theorem \ref{thm:lollipop}. Hence, the algorithm is optimal.

\section{Conclusion}\label{section:conclusions}
In this work, we proved optimal spatial search occurs on cliques even in the presence of an infinite path. 
This generalized a known result of Farhi and Gutmann \cite{fg98} to the infinite setting.
We view this as a first step in showing that optimal spatial search is robust against 
an adversary modeled as an infinite-dimensional external quantum probe. 
Interesting directions for future work include extending the result to multiple tails or 
to tails induced by more general Jacobi matrices and strengthening the lower bound to
other families of infinite graphs.

Our work was motivated by a question posed in \cite{btvx}. 
We mention that discrete-time quantum walk on graphs with tails was studied in \cite{kss}.

\section*{Acknowledgments}
Work started during the workshop ``Graph Theory, Algebraic Combinatorics, and Mathematical Physics''
at Centre de Recherches Math\'{e}matiques (CRM), Universit\'{e} de Montr\'{e}al.
C.T. would like to thank CRM for its hospitality and support during his sabbatical visit.
W.X.\ was supported by NSF grant DMS-2212755.
We thank Pierre-Antoine Bernard and Luc Vinet for discussions.

\bibliography{qwalk}
\bibliographystyle{plain}

\end{document}